\documentclass[preprint,aps,tightenlines,floatfix,showpacs]{revtex4}
\usepackage{graphics}
\usepackage{bm}
\usepackage{amsmath}
\usepackage{amssymb}

\begin{document}
  \begin{center}
    {\large\bf Scattering as a tool to study nuclear structure toward
    the drip lines} \\ \vskip 5 mm
    S. Karataglidis$^{(a)}$ and K. Amos$^{(b)}$ \\
    $^{(a)}$ Department of Physics and Electronics, Rhodes University
    Grahamstown, 6140, South Africa \\
    $^{(b)}$ School of Physics, University of Melbourne, Victoria, 3010,
    Australia
  \end{center}
  \begin{center}
    {\large\bf  Abstract}
  \end{center}
  Results are  presented for the  elastic scattering of  electrons and
  protons  from  the  exotic  He  and  Li  isotopes.  Comparison  with
  scattering results  from the stable He  and Li nuclei  allows for an
  investigation   into  the   effects  that   the   extensive  neutron
  distributions  have on the  charge density.  For comparison  we also
  consider  the  proton  halo  nucleus  $^8$B.  The  consequences  and
  possible suggestions for proposed electron scattering facilities for
  exotic nuclei are discussed.

  \section{Introduction}

  Current microscopic understanding  of structures of exotic nuclei, both
  of  halo and  skin type,  has  been made  possible by  their use  in
  analysis of data from the  elastic and inelastic scattering of those
  nuclei  from hydrogen.  In inverse  kinematics such  data  equate to
  elastic and inelastic proton scattering  and so one is able to probe
  the matter densities  of such systems. As the  $pn$ component of the
  nucleon-nucleon  ($NN$) force is  dominant at  intermediate energies
  \cite{Am00}, proton scattering  primarily probes the neutron density
  and  vice-versa. As  an  example,  the neutron  halo  in $^6$He  was
  clearly  established  by  analyses  of  $^6$He-$p$  scattering  data
  \cite{La01,St02}. To do so requires use of predictive models
  of  nucleon-nucleus ($NA$)  scattering  which are  sensitive to  the
  details  of the  matter  density.  One such is the Melbourne model
  \cite{Am00} which has  been successful in predicting observables for 
  many elastic and inelastic $NA$ scatterings.

  But  analyses  of  $pA$  scattering data  from  exotic  neutron-rich
  systems generally do not allow for the direct investigation of their
  proton  densities. With the  development of  electron-ion colliders,
  and   the  more  novel   approach  of   constructing  Self-Confining
  Radioactive  Ion Targets  (SCRIT),  the possibility now exists  for 
  the  direct investigation of the  charge density of exotic halo   and 
  skin nuclei  by electron  scattering. (For both developments, see the
  review  by Suda and  Wakasugi \cite{Su05}.)  Additionally, kinematic
  reconstruction   of  events  from  those facilities   will  provide
  measurements of both longitudinal  and transverse form factors. Such
  will  probe both  the charge  and  current densities  of the  exotic
  nuclei,  complementing  the  information  on  the  matter  densities
  provided by modern $NA$ scattering analyses \cite{Am00}.

  Taken together,  a complete map  of exotic nuclei will  be possible.
  The purposes of  this paper is to consider  both proton and electron
  scattering from  exotic nuclei to  identify effects of the  halo, or
  skin,  in  data from  complementary  experiments. Further,  by
  comparing  also to  analyses of  data  from stable  nuclei, one  may
  consider how  the densities change  with the addition of  protons or
  neutrons  as   one  approaches  the  drip  lines.

  In the next three sections, outlines are given of the microscopic 
  models we use
  to evaluate  proton  and electron scattering  observables. Then,
  results are  presented for scattering  from the He and  Li isotopes,
  concluding with $^8$B, after which follow our conclusions.

  \section{Intermediate energy nucleon scattering}
 
  A prerequisite for the use  of proton scattering data as an analysis
  tool  for nuclear  structure is  a model  which is  able  to predict
  scattering observables without any \textit{a posteriori} adjustments
  of parameters. Such a model  is that using the Melbourne force, when
  all nuclear  structure and scattering details are  preset. With this
  prescription, elastic scattering differential cross section and spin
  observable  data  have  been  well  reproduced. The  full  model  is
  described  in  a review  \cite{Am00}  from  which  a brief  overview
  follows.

  To describe intermediate energy $NA$ data one begins with a credible
  effective $NN$  force. In general,  such should link to the $NN$  
  $g$ matrices, which we take to be solutions  of the 
  Brueckner-Bethe-Goldstone (BBG) equations for infinite nuclear matter,
  \textit{viz.}
  \begin{equation}
    g\left( \mathbf{q}', \mathbf{q}; K \right) = V\left( \mathbf{q}',
    \mathbf{q} \right) + \int V\left( \mathbf{q}', \mathbf{k}' \right)
    \frac{ Q\left( \mathbf{k}', \mathbf{K}; k_f \right) }{ \left[
    E\left( \mathbf{k}, \mathbf{K} \right) - E\left( \mathbf{k}',
    \mathbf{K} \right) \right] } \, g\left( \mathbf{k}', \mathbf{q};
    \mathbf{K} \right) \, d\mathbf{k}',
  \end{equation}
  where  $Q$ is  a Pauli  blocking  operator, and  medium effects  are
  incorporated  into  the  auxiliary  potentials entering  the  energy
  denominator  \cite{Am00}. The center-of-mass  and Fermi  momenta are
  denoted by $K$ and $k_f$, respectively.

   The BBG $g$ matrices are mapped to a usable coordinate space form of 
  complex, energy and density dependent, model $NN$ effective interaction 
  ($g_{\text{eff}}$), which when folded with any microscopic model structure of
  the target gives the $NA$ optical potential of the form,
  \begin{align}
    U\left( \mathbf{r}, \mathbf{r}'; E \right) & = \delta\left(
    \mathbf{r} - \mathbf{r}' \right) \sum_i n_i \int
    \varphi^{\ast}_{i}(\mathbf{s}) g_D\left( \mathbf{r}, \mathbf{s}; E
    \right) \varphi_i(\mathbf{s}) \, d\mathbf{s} + \sum_i n_i \varphi^{\ast}_i(\mathbf{r})
    g_E\left( \mathbf{r}, \mathbf{r}'; E \right) \varphi_i(
    \mathbf{r}') \nonumber \\
    & = U_D\left( \mathbf{r}; E \right) \delta\left( \mathbf{r} -
    \mathbf{r}' \right) + U_E\left( \mathbf{r}, \mathbf{r}'; E
    \right),
  \end{align}
  where the subscripts  $D,E$ denote the direct and  exchange parts of
  the  optical potential, respectively.  The sums  are taken  over the
  bound  state   single-particle  orbits  for  which   $n_i$  are  the
  associated occupation  numbers. In evaluations with  the DWBA98 code
  \cite{Ra98},  the single-particle  wave  functions may  be taken  as
  either of harmonic oscillator (HO) or Woods-Saxon (WS) form. A variant
  code now has been developed to use Skyrme-Hartree-Fock wave functions.
  The resultant complex, energy and  density-dependent  ($g$-folding) 
  optical  potential so formed, contains central and  spin-orbit terms.  
  With such generated in the DWBA98 code \cite{Ra98}, they are then used
  to calculate differential  cross sections  and  spin-dependent
  observables. They are also used in that code to evaluate cross sections
  and the like for inelastic scattering with the self same $g_{\text{eff}}$ 
being 
  taken as the transition operator.  That inelastic  scattering   is  
  calculated   in   the  distorted   wave approximation (DWA).
  For this the transition amplitude may be written as
  \begin{equation}
    T^{M_f M_i \nu' \nu}_{J_fJ_i}(\theta) = \left\langle
    \chi^{(-)}_{\nu'} \right| \left\langle \Psi_{J_fM_f} \right| A
    g_{\text{eff}}(0,1) \mathcal{A}_{01} \left\{ \left|
    \chi^{(+)}_{\nu} \right\rangle \left| \Psi_{J_i M_i} \right\rangle
    \right\},
  \end{equation}
  where  $\chi^{(\pm)}$ denoted  the distorted  wave functions  for an
  incoming/outgoing   nucleon,  respectively,   and  0,1   denote  the
  coordinates of the projectile and bound state nucleon, respectively.
  A  co-factor expansion  of the  many-body bound-state  wave function
  \cite{Am00} leads to,
  \begin{multline}
    T^{M_f M_i \nu' \nu}_{J_fJ_i}(\theta) = \sum_{\alpha_1 \alpha_2
    m_1 m_2} \sum_{JM} \frac{(-1)^{j_1 - m_1}}{\sqrt{2J_f + 1}} \left \langle
    \left. j_2 \, m_2 \, j_1 \, -m_1 \right| J_f \, M_f \right\rangle
    \left\langle \left. J_i \, M_i \, J \, M\, \right| J_f \, M_f
    \right\rangle
    \\
    \times \left\langle J_f \left\| \left[ a^{\dag}_{\alpha_2} \times
      \tilde{a}_{\alpha_1} \right]^J \right\| J_i \right\rangle \\
    \times \left\langle \chi^{(-)}_{\nu'}(0) \right| \left\langle
    \varphi_{\alpha_2}(1) \right| A g_{\text{eff}}(0,1) \mathcal{A}_{01}
    \left\{ \left| \chi^{(+)}_{\nu}(0) \right\rangle \left|
    \varphi_{\alpha_1}(1) \right\rangle \right\}
  \end{multline}
  for  an angular momentum  transfer $J$.  In this  equation, $\alpha$
  designates the set  of single-particle quantum numbers $\{  n, l, j,
  m_{\tau}\}$, where $\tau$ is the nucleon isospin.
  
  \section{Electron scattering}

  The  approach  we adopt  for  the  calculation  of longitudinal  and
  transverse  form  factors  follows  that  of  deForest  and  Walecka
  \cite{Fo66},  and of  Karataglidis, Halse,  and Amos  \cite{Ka95}. A
  brief summary is given herein.
  
  Form factors  for electron  scattering between nuclear  states $J_i$
  and $J_f$  involving angular momentum transfer $J$  may be expressed
  as
  \begin{equation}
    \left| F^\eta_J(q) \right|^2 = \frac{1}{2J_i+1} \left(
    \frac{4\pi}{Z^2} \right) \left| \left\langle J_f \left\|
    T^\eta_J(q) \right\| J_i \right\rangle \right|^2,
  \end{equation}
  where  $\eta$  selects  the  longitudinal, transverse  electric,  or
  transverse magnetic form  factors. Assuming one-body operators, the
  reduced matrix elements may be expressed in the form
  \begin{equation}
    \left\langle J_f \left\| T^{\eta}_J(q) \right\| J_i \right\rangle
    = \text{Tr}\left( SM \right)/\sqrt{2J+1},
  \end{equation}
  where $S$  is the  matrix of one-body  transition densities  and $M$
  contains  the  single-particle   matrix  elements  of  the  one-body
  longitudinal or transverse electromagnetic operators. Bare operators
  are   used  for   the   results  presented   herein,  and   explicit
  meson-exchange-current (MEC) effects  are ignored. However, MEC have
  been incorporated implicitly in the transverse electric form factors
  in the long-wavelength limit by using Siegert's theorem \cite{Fr85}.

  \section{Nuclear Structure}
  The  nuclear  structure information  entering  in  analyses of  both
  nucleon and electron scattering data are the one-body density matrix
  elements (OBDME), $S_{\alpha_1 \alpha_2 J}$ \textit{viz.}
  \begin{equation}
    S_{\alpha_1 \alpha_2 J} = \left\langle J_f \left\| \left[
    a^{\dag}_{\alpha_2} \times \tilde{a}_{\alpha_1} \right]^J \right\|
    J_i \right\rangle.
  \end{equation}
  For  elastic  scattering these  reduce  essentially  to the  nucleon
  occupation numbers, and we have used the shell model to obtain them.

  One also  requires specification of  the single-particle bound-state
  wave  functions,  for  which  we  have  used both  HO  and  WS  wave
  functions. In  the case of  scattering from halo nuclei,  the former
  gives the ``nonhalo''  results while results from use  of the latter
  are denoted ``halo'' when the binding energies of the valence orbits
  are set to the single-particle separation energy.
    
  \section{Results}

\begin{figure}
    \scalebox{0.4}{\includegraphics*{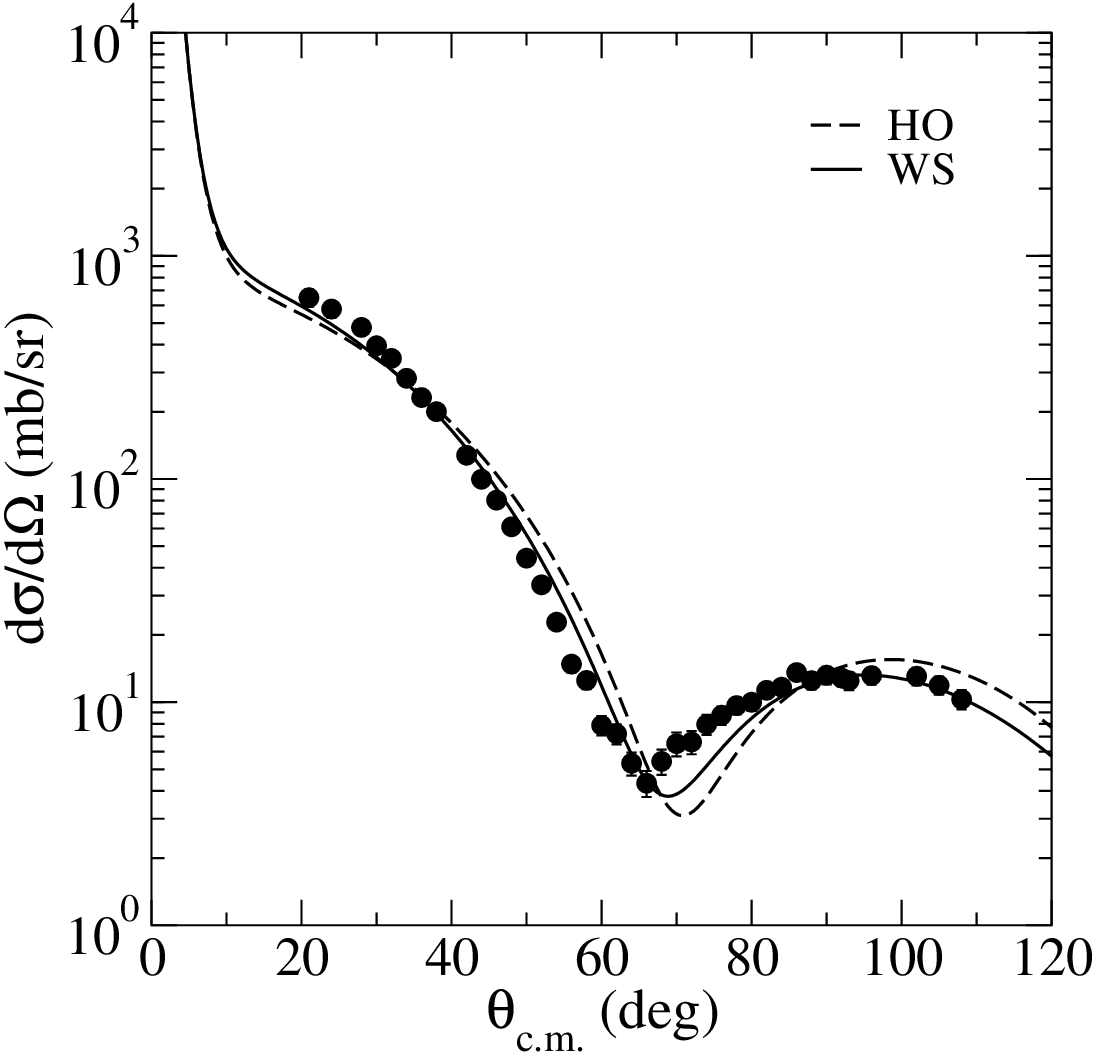}}
    \caption{\label{he8comp}   Comparison  of   predictions   for  the
    differential cross section for the elastic scattering of $16A$~MeV
    $^8$He ions  from hydrogen.  The data are  those of Skaza  {\em et
    al.} \cite{Sk05}.}
  \end{figure}

  We first  consider elastic and inelastic  intermediate energy proton
  scattering. It has been established in $^6$He-hydrogen scattering at
  $41A$~MeV \cite{La01} that  $^6$He has a neutron halo,  and this was
  subsequently  confirmed  in  analyses  of data  taken  at  $25A$~MeV
  \cite{St02}. Those were made using  the WS functions defined in Ref.
  \cite{Ka00}  for  the  specification  of  the  halo.  New  data  are
  available  for the  elastic scattering  of $^8$He  from  hydrogen at
  $16A$~MeV \cite{Sk05}.  A complete $(0+2+4)\hbar\omega$  shell model
  with the Zheng $G$ matrix  interaction \cite{Zh95} was used with the
  WS functions of Ref. \cite{Ka00} but without the shallow binding for
  the halo neutrons in $^6$He. Comparison of our prediction with those
  data is presented in Fig.~\ref{he8comp}.
  The  similarity between  the results  obtained using  the HO  and WS
  single-particle wave functions confirms  that $^8$He does not have a
  neutron halo  but rather it has  a neutron skin. Our  results may be
  compared with  those presented  in Ref.~\cite{Sk05}, which  used the
  JLM effective interaction together  with an explicit coupling to the
  $(p,d)$  channel   with  $^7$He  as  an   intermediate  state.  That
  particular model is  most problematic: $^7$He may not  be treated as
  an  intermediate state  in  elastic scattering  as  that nucleus  is
  particle unstable.  It will break up immediately  ($\tau_{1/2} = 2.2
  \times  10^{-21}$~s  \cite{TUNL})  and  so the  probability  of  the
  reformation  of $^8$He from  $^7$He$+d$ is  negligible. Use  of that
  model required an unreasonably low normalization ($\lambda_W = 0.2$)
  of the imaginary part of the optical potential \cite{Sk05} to obtain
  agreement with the data.

\begin{figure}
    \scalebox{0.4}{\includegraphics*{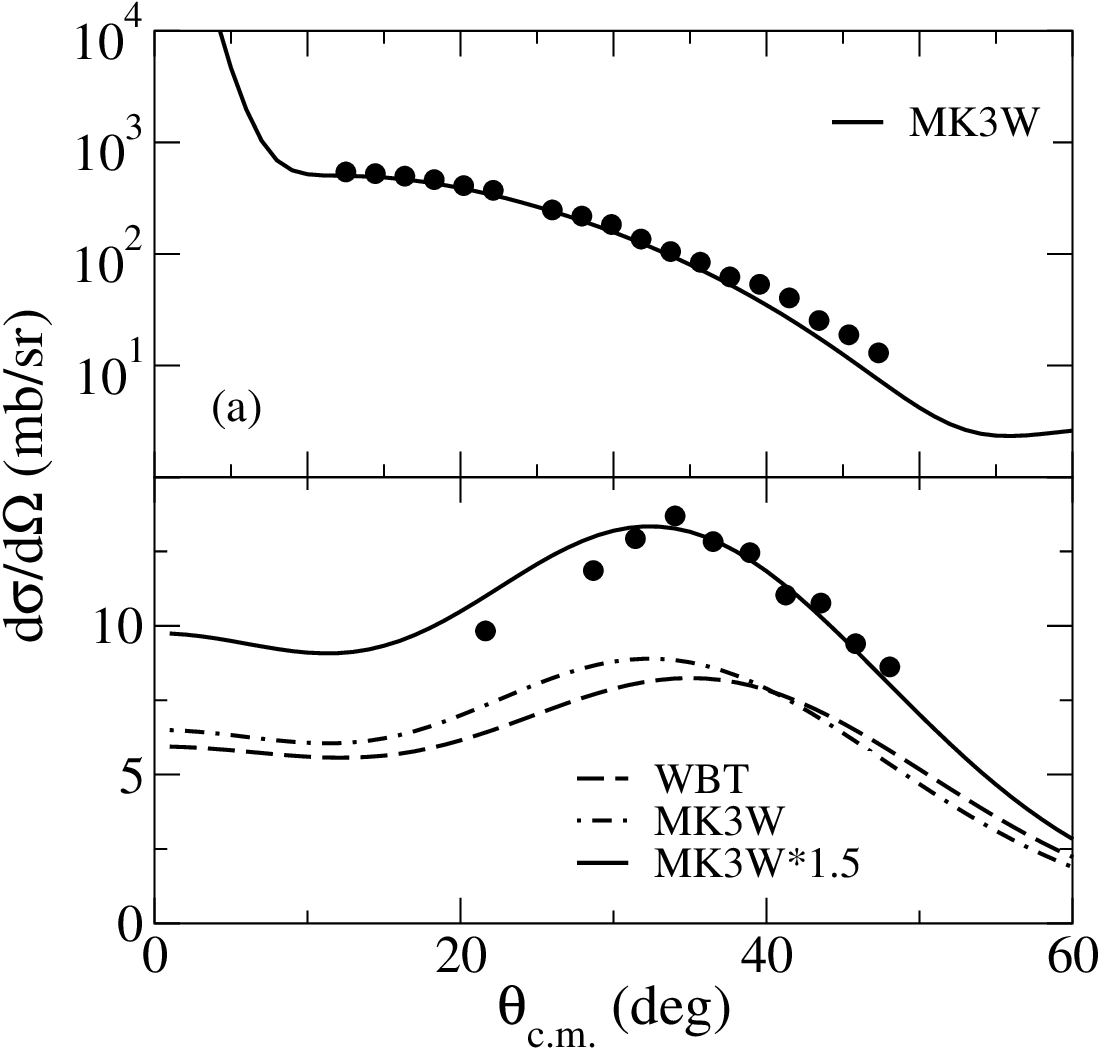}}
    \scalebox{0.4}{\includegraphics*{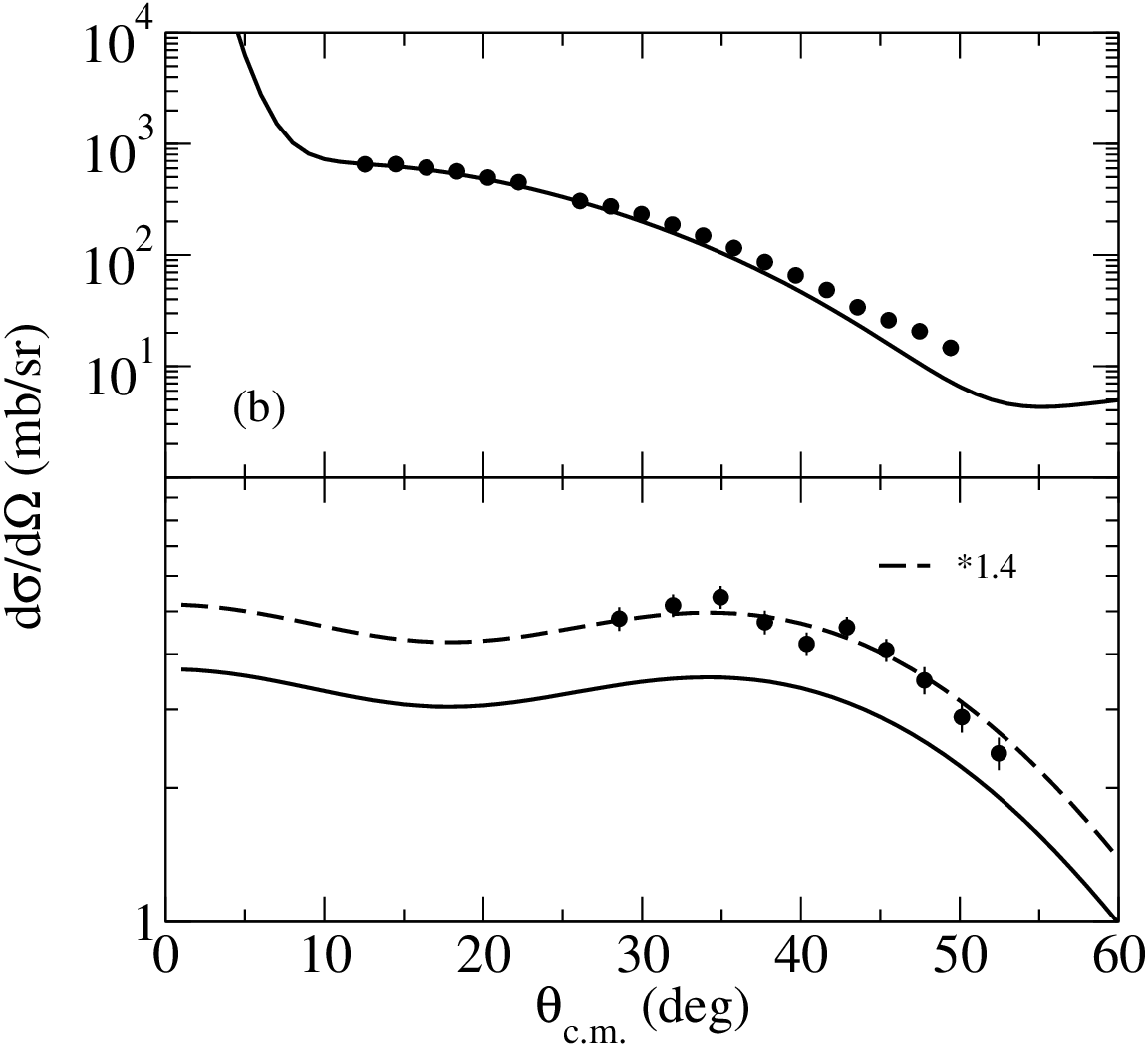}}
    \caption{\label{c10-c11} $^{10}$C (a)  and $^{11}$C (b) scattering
    from hydrogen  at $45A$ and $40A$~MeV, respectively.  The top part
    of the panels show the  elastic scattering while the bottom panels
    show the inelastic  scattering as described in the  text. The data
    are from Ref. \cite{Jo05}.}
  \end{figure}

  Data  have also  been obtained  recently for  elastic  and inelastic
  scattering of  $^{10}$C and  $^{11}$C from hydrogen  \cite{Jo05}, at
  $45A$ and $40A$ MeV, respectively. We have analyzed those data using
  densities  obtained from a  complete $(0+2)\hbar\omega$  shell model
  using  the MK3W \cite{Wa89} and WBT \cite{Wa92} interactions.  
 Our results  are compared
  with    the   available    data   in    Figs.~\ref{c10-c11}(a)   and
  \ref{c10-c11}(b).
  HO wave functions with $b = 1.6$~fm were used to obtain both sets of
  results. The  data for the  elastic scattering from both  nuclei are
  well  reproduced, indicating  that the  structure model  suffices to
  specify  the ground  state densities  of $^{10,11}$C. 
  The result for elastic scattering of ${}^{10}$C using the WBT 
  interaction is not shown as it is very similar to that found using 
  the MK3W.  Note  that no
  adjustment of the input potentials  or wave functions have been made
  in  obtaining  these  results.   As  with  the  case  of  $^8$He$+p$
  scattering,  this is  in contrast  to the  earlier  JLM calculations
  which required  a renormalization of  the \textit{real} part  of the
  potential  by $\sim  10$\%,  an  adjustment for  which  there is  no
  \textit{a  priori} reason,  other than  it appears  to be  a general
  trend.  This is  more  a  serious problem  with  the underlying  JLM
  potential.  A more  drastic  renormalization is  required with  that
  model to  describe the data  for the elastic scattering  from $^6$He
  \cite{Jo05} as well.

  The data for the inelastic scattering to the $2^+$ state in $^{10}$C
  and $\frac{5}{2}^-$ state in $^{11}$C are shown in the bottom panels
  of  Figs.~\ref{c10-c11}(a)  and  \ref{c10-c11}(b), respectively.  In
  each case,  the results of our calculations  underestimate the data,
  both requiring a renormalization  of around 1.4 to obtain agreement.
  But in both cases the  shape is reproduced very well. This indicates
  that  some additional  core polarization  terms are  missing  in the
  structure model. Higher-order  $\hbar\omega$ components in the model
  space are  needed as studies of  $p-{}^{12}$C scattering \cite{Am00}
  showed  them to  be for  evaluation  of the  $2^+$ (4.43~MeV)  state
  excitation.

\begin{figure}
    \scalebox{0.4}{\includegraphics*{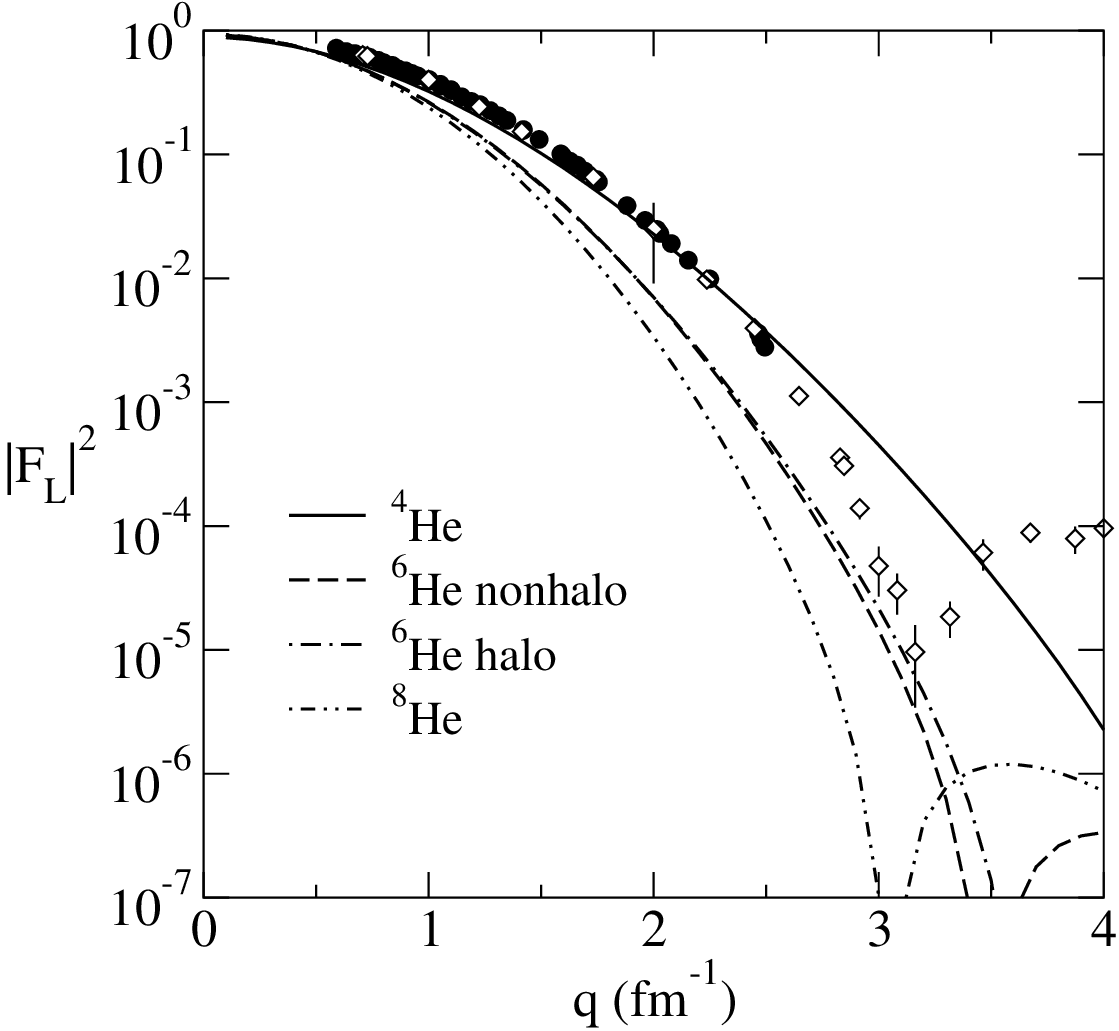}}
    \caption{\label{he_elec} Elastic electron scattering form factors
    for $^{4,6,8}$He. The data for the $^4$He form factor are those of
    McCarthy {\em et al.} \cite{Mc77}.}
  \end{figure}

  The  longitudinal  elastic  electron  scattering  form  factors  for
  $^{4,6,8}$He are displayed in Fig.~\ref{he_elec}.
  The comparison of the $^4$He  form factor with the available data is
  quite  good  up  to  2.5~fm$^{-1}$.  This  is  consistent  with  the
  predicted  charge radius  of 1.71~fm,  as compared  to  the measured
  value  of $1.671  \pm 0.014$~fm  \cite{Ot85}. The  form  factors for
  $^6$He and $^8$He have been calculated using WS functions as used in
  the analysis of proton scattering.  The addition of neutrons to form
  $^6$He  and $^8$He pull  the charge  density out  and thus  the form
  factors  decrease with  momentum  transfer. Note  that the  detailed
  properties  of the neutron  halo in  $^6$He do  not affect  the form
  factor; it is only the presence  of the extra 2 neutrons that causes
  the change to the proton distribution.

\begin{figure}
    \scalebox{0.35}{\includegraphics*{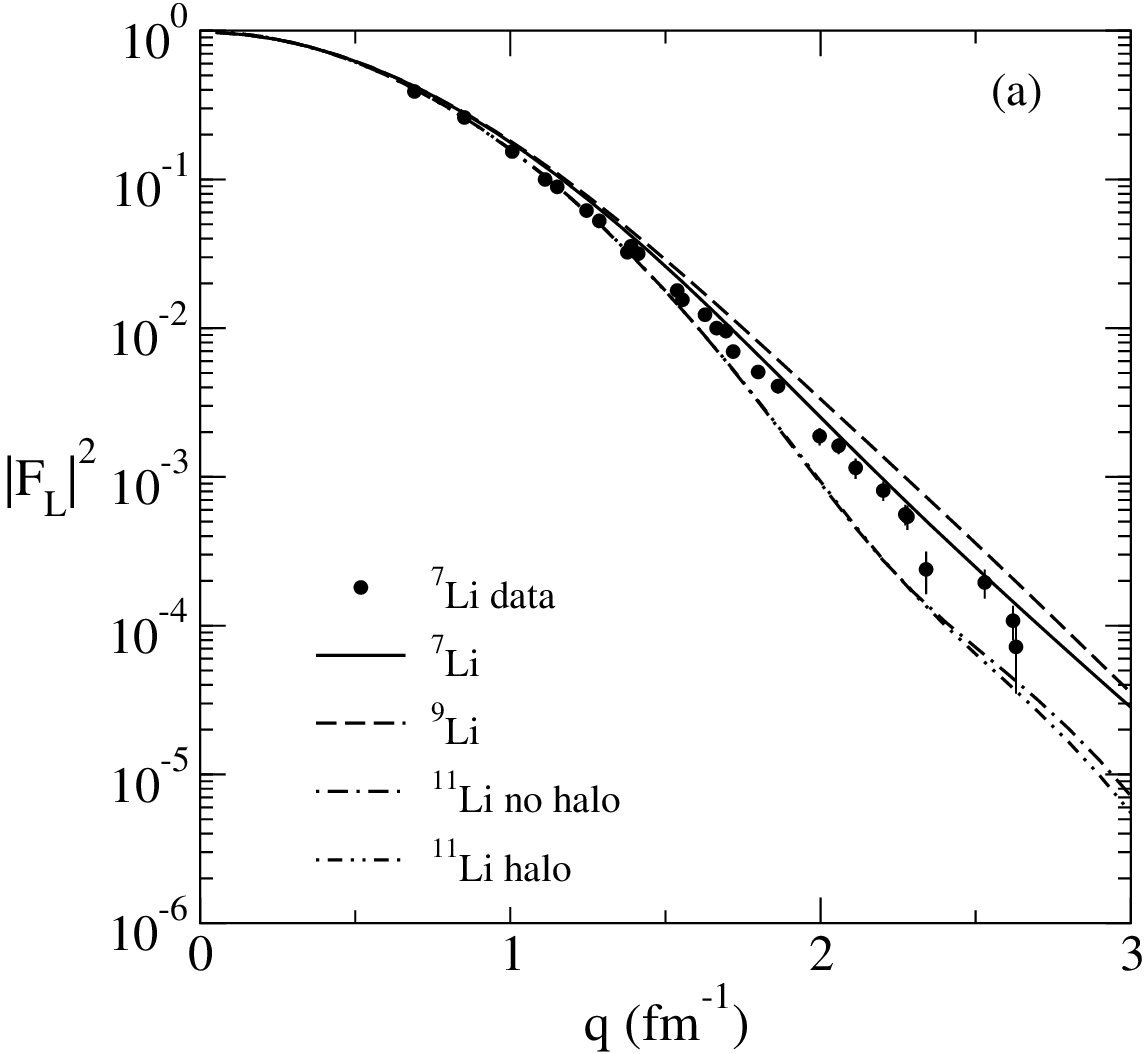}}
    \scalebox{0.36}{\includegraphics*{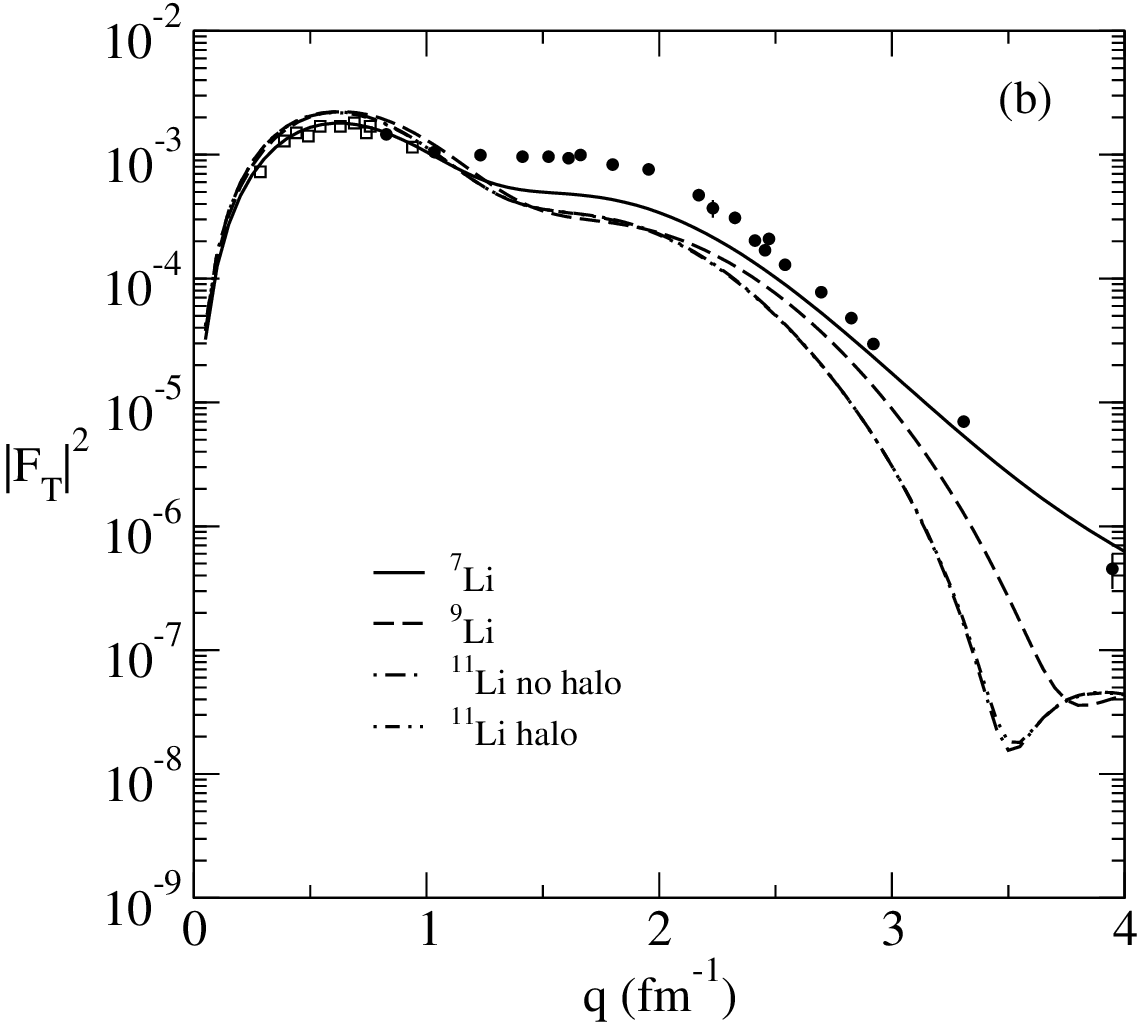}}
    \caption{\label{li_elec} Longitudinal (a) and transverse (b) form
      factors for $^7$Li, $^9$Li, and $^{11}$Li. The data are from
      Refs. \cite{Su67,Li89,Ni71}.}
  \end{figure}

  The  longitudinal and  transverse elastic  electron  scattering form
  factors  for   $^7$Li,  $^9$Li   and  $^{11}$Li  are   displayed  in
  Figs.~\ref{li_elec}(a) and \ref{li_elec}(b), respectively.
  The level of  agreement between the results of  our calculations for
  $^7$Li with  data \cite{Su67,Li89,Ni71} is quite  good. The addition
  of two neutrons  does not change the form  factors substantially and
  so the  charge density  for $^9$Li is  little changed from  that for
  $^7$Li. But a noticeable change  is observed in the form factors for
  $^{11}$Li, evidenced by the longitudinal form factor decreasing with
  momentum transfer. As  with $^6$He, this change does  not come about
  with the halo  specifics in this nucleus; rather, it  is due only to
  the coupling of the 4 extra neutrons to the $^7$Li core.

\begin{figure}
    \scalebox{0.4}{\includegraphics*{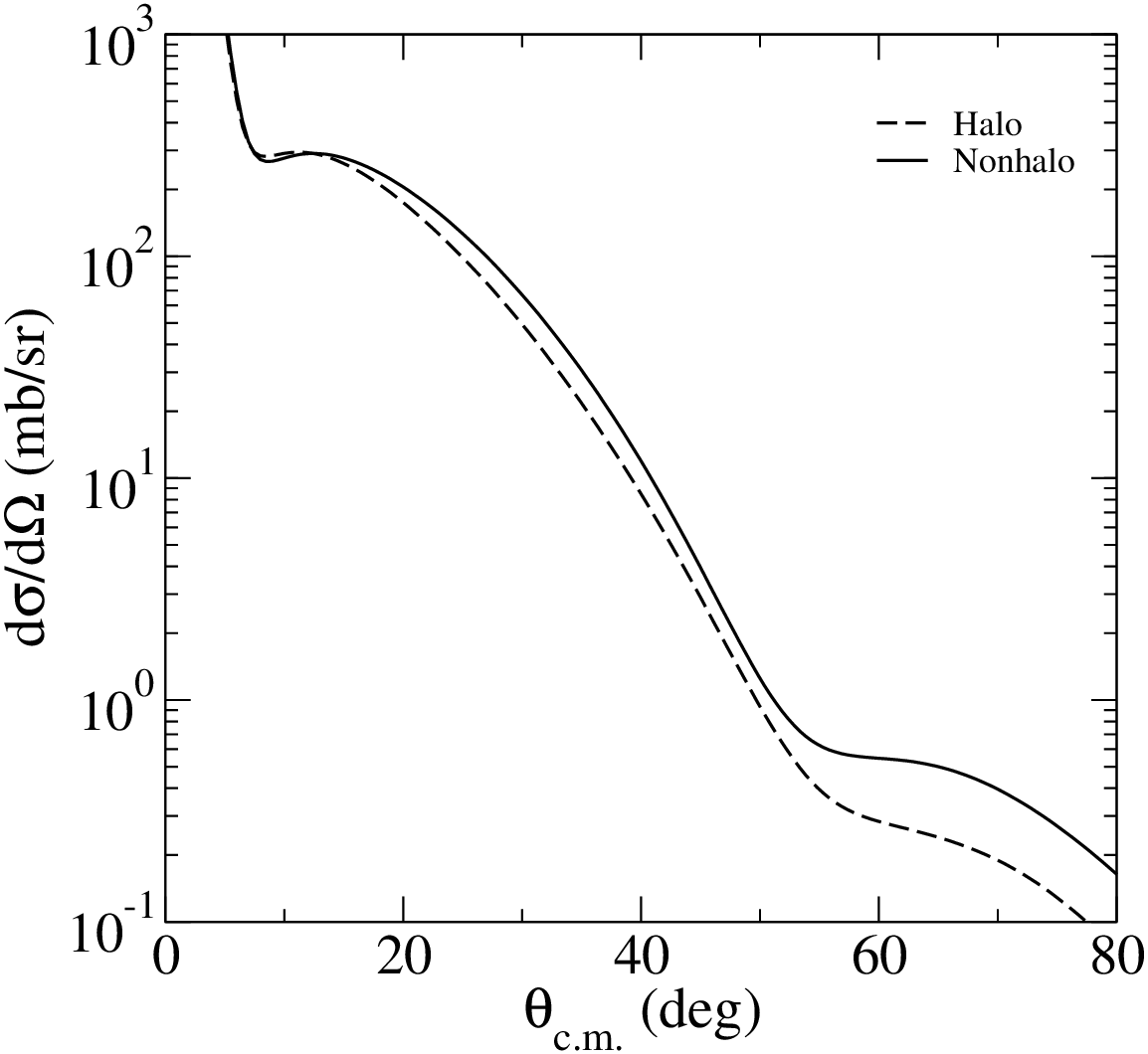}}
    \caption{\label{b8_proton} Differential cross section predictions
      for the elastic scattering of 65 MeV protons from $^8$B.}
  \end{figure}

  As a comparison  to the neutron halos, we now  turn our attention to
  the scattering from a proton  halo nucleus, namely $^8$B. A complete
  $(0+2+4)\hbar\omega$  shell   model  using  the   Zheng  interaction
  \cite{Zh95}  was  used  to  obtain  the  OBDME.  The  single  proton
  separation energy from $^8$B is  137 keV \cite{TUNL04} and we use WS
  functions with that binding energy to specify the halo. The non-halo
  specification uses the  same set of WS functions  as for $^8$He. Our
  prediction  of  the  differential  cross  section  for  the  elastic
  scattering   of   65   MeV   protons   from  $^8$B   is   shown   in
  Fig.~\ref{b8_proton}.
  As  for  $^6$He  and  $^{11}$Li,  the halo  manifests  itself  as  a
  reduction in  the cross section  at large angles.  This is due  to a
  reduction  of the  proton  density  at the  core,  a consequence  of
  requiring the density to extend out to large radius while preserving
  particle number.

\begin{figure}
    \scalebox{0.365}{\includegraphics*{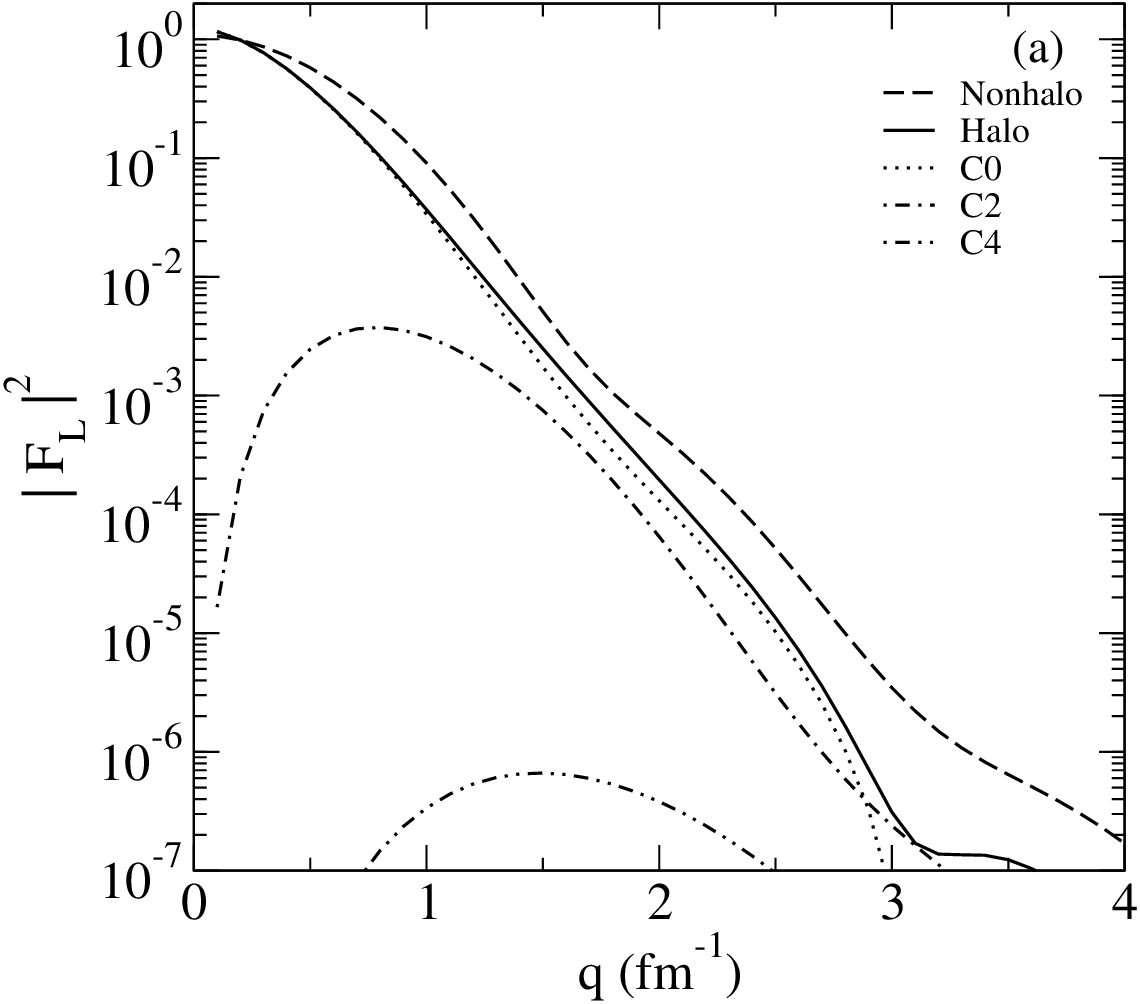}}
    \scalebox{0.35}{\includegraphics*{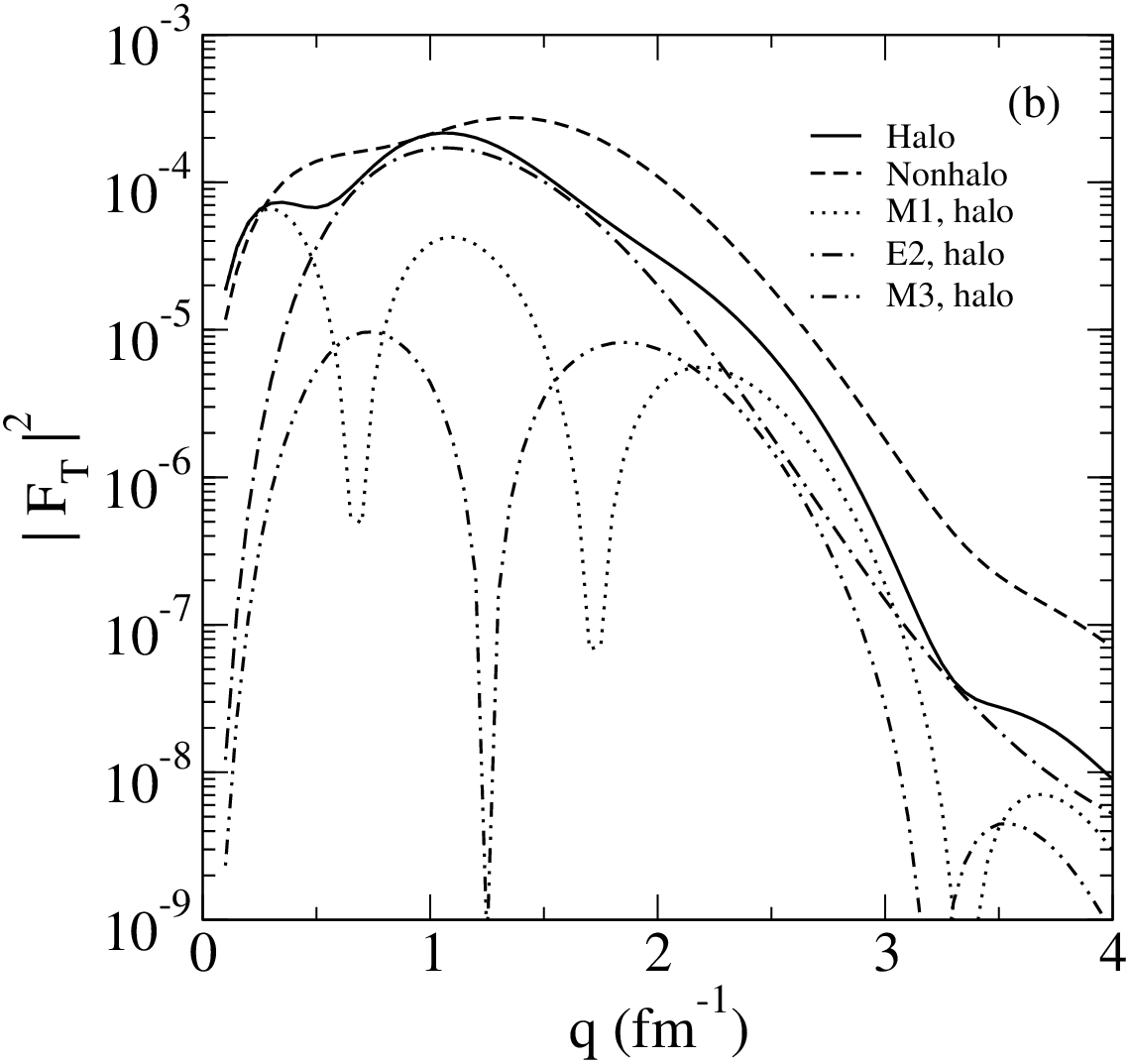}}
    \caption{\label{b8_elec} Longitudinal (a) and transverse (b)
      elastic electron scattering form factors for $^8$B.}
  \end{figure}

  Our predictions for the longitudinal and transverse elastic electron
  scattering  form  factors are  shown  in Figs.~\ref{b8_elec}(a)  and
  \ref{b8_elec}(b).
  The  longitudinal form  factor is  dominated by  the  $C0$ component
  below  1~fm$^{-1}$ while the  $C2$ contribution  becomes significant
  above  that  value  of  momentum  transfer. The  $C4$  component  is
  negligible.  The  effects of  the  proton  halo,  in this  case,  is
  distinct. Notably, the form  factor decreases with momentum transfer
  which is consistent with  a more extensive charge distribution. That
  effect   is  also   noticeable  in   the  transverse   form  factor,
  Fig.~\ref{b8_elec}(b),  which is  dominated by  the  $E2$ component.
  Thus, the halo  also affects the current density  as that is related
  to the charge density through the continuity equation.
  
  \section{Conclusions}

  We have presented results for  the elastic scattering of protons and
  electrons from  exotic nuclei, including  halo systems, in  order to
  elicit details of the effects  of the extended nucleon matter on the
  charge density. While proton  scattering is sensitive to the details
  of the neutron halo, it has been shown, in the longitudinal electron
  scattering form  factors of $^6$He  and $^{11}$Li, that  the neutron
  halo does not affect the  charge density. Instead the charge density
  is extended naturally  only through the addition of  neutrons to the
  stable isotopes. That will  occur irrespective of whether the exotic
  nucleus is a neutron halo or  a skin. The transverse form factors of
  the neutron-rich systems portray similar effects where the change to
  the current  density is due simply  to the addition  of neutrons and
  not specifically if they form a neutron halo or skin.

  The  situation   for  proton   halos  is  entirely   different.  The
  longitudinal  and transverse  electron scattering  form  factors for
  $^8$B  are  significantly  reduced.  This  is  consistent  with  the
  extension  of  the  charge  density  due to  the  proton  halo.  The
  transverse form factor  is dominated by the $E2$  term which is also
  affected by the proton halo.

  With the introduction of the electron-ion collider, it is hoped that
  investigations of the proton halos will be possible. Transverse form
  factors should  also be measured  as such measurements will  also be
  possible  in  the  colliders.  Analysis  with  complementary  proton
  scattering  data will  allow  for the  possibility  of the  complete
  mapping the  densities of exotic  nuclei allowing for  more detailed
  microscopic tests of structure models in use.
  
  \bibliography{var_proc}

\begin{thebibliography}{21}
\expandafter\ifx\csname natexlab\endcsname\relax\def\natexlab#1{#1}\fi
\expandafter\ifx\csname bibnamefont\endcsname\relax
  \def\bibnamefont#1{#1}\fi
\expandafter\ifx\csname bibfnamefont\endcsname\relax
  \def\bibfnamefont#1{#1}\fi
\expandafter\ifx\csname citenamefont\endcsname\relax
  \def\citenamefont#1{#1}\fi
\expandafter\ifx\csname url\endcsname\relax
  \def\url#1{\texttt{#1}}\fi
\expandafter\ifx\csname urlprefix\endcsname\relax\def\urlprefix{URL }\fi
\providecommand{\bibinfo}[2]{#2}
\providecommand{\eprint}[2][]{\url{#2}}

\bibitem[{\citenamefont{Amos et~al.}(2000)\citenamefont{Amos, Dortmans, von
  Geramb, Karataglidis, and Raynal}}]{Am00}
\bibinfo{author}{\bibfnamefont{K.}~\bibnamefont{Amos}},
  \bibinfo{author}{\bibfnamefont{P.~J.} \bibnamefont{Dortmans}},
  \bibinfo{author}{\bibfnamefont{H.~V.} \bibnamefont{von Geramb}},
  \bibinfo{author}{\bibfnamefont{S.}~\bibnamefont{Karataglidis}},
  \bibnamefont{and} \bibinfo{author}{\bibfnamefont{J.}~\bibnamefont{Raynal}},
  \bibinfo{journal}{Adv. in Nucl. Phys.} \textbf{\bibinfo{volume}{25}},
  \bibinfo{pages}{275} (\bibinfo{year}{2000}), \bibinfo{note}{and referencees
  cited therein.}

\bibitem[{\citenamefont{Lagoyannis et~al.}(2001)}]{La01}
\bibinfo{author}{\bibfnamefont{A.}~\bibnamefont{Lagoyannis}}
  \bibnamefont{et~al.}, \bibinfo{journal}{Phys. Lett.}
  \textbf{\bibinfo{volume}{B518}}, \bibinfo{pages}{27} (\bibinfo{year}{2001}).

\bibitem[{\citenamefont{Stepantsov et~al.}(2002)}]{St02}
\bibinfo{author}{\bibfnamefont{S.}~\bibnamefont{Stepantsov}}
  \bibnamefont{et~al.}, \bibinfo{journal}{Phys. Lett.}
  \textbf{\bibinfo{volume}{542B}}, \bibinfo{pages}{35} (\bibinfo{year}{2002}).

\bibitem[{\citenamefont{Suda and Wakasugi}(2005)}]{Su05}
\bibinfo{author}{\bibfnamefont{T.}~\bibnamefont{Suda}} \bibnamefont{and}
  \bibinfo{author}{\bibfnamefont{M.}~\bibnamefont{Wakasugi}},
  \bibinfo{journal}{Prog. Part. Nucl. Phys.} \textbf{\bibinfo{volume}{55}},
  \bibinfo{pages}{417} (\bibinfo{year}{2005}).

\bibitem[{\citenamefont{Raynal}(1998)}]{Ra98}
\bibinfo{author}{\bibfnamefont{J.}~\bibnamefont{Raynal}},
  \emph{\bibinfo{title}{Computer Program DWBA98, NEA 1209/05}}
  (\bibinfo{year}{1998}).

\bibitem[{\citenamefont{deForest and Walecka}(1966)}]{Fo66}
\bibinfo{author}{\bibfnamefont{T.}~\bibnamefont{deForest}} \bibnamefont{and}
  \bibinfo{author}{\bibfnamefont{J.~D.} \bibnamefont{Walecka}},
  \bibinfo{journal}{Adv. Phys.} \textbf{\bibinfo{volume}{15}},
  \bibinfo{pages}{1} (\bibinfo{year}{1966}).

\bibitem[{\citenamefont{Karataglidis et~al.}(1995)\citenamefont{Karataglidis,
  Halse, and Amos}}]{Ka95}
\bibinfo{author}{\bibfnamefont{S.}~\bibnamefont{Karataglidis}},
  \bibinfo{author}{\bibfnamefont{P.}~\bibnamefont{Halse}}, \bibnamefont{and}
  \bibinfo{author}{\bibfnamefont{K.}~\bibnamefont{Amos}},
  \bibinfo{journal}{Phys. Rev. C} \textbf{\bibinfo{volume}{51}},
  \bibinfo{pages}{2494} (\bibinfo{year}{1995}).

\bibitem[{\citenamefont{Friar and Haxton}(1985)}]{Fr85}
\bibinfo{author}{\bibfnamefont{J.~L.} \bibnamefont{Friar}} \bibnamefont{and}
  \bibinfo{author}{\bibfnamefont{W.~C.} \bibnamefont{Haxton}},
  \bibinfo{journal}{Phys. Rev. C} \textbf{\bibinfo{volume}{31}},
  \bibinfo{pages}{2027} (\bibinfo{year}{1985}).

\bibitem[{\citenamefont{Karataglidis et~al.}(2000)\citenamefont{Karataglidis,
  Dortmans, Amos, and Bennhold}}]{Ka00}
\bibinfo{author}{\bibfnamefont{S.}~\bibnamefont{Karataglidis}},
  \bibinfo{author}{\bibfnamefont{P.~J.} \bibnamefont{Dortmans}},
  \bibinfo{author}{\bibfnamefont{K.}~\bibnamefont{Amos}}, \bibnamefont{and}
  \bibinfo{author}{\bibfnamefont{C.}~\bibnamefont{Bennhold}},
  \bibinfo{journal}{Phys. Rev. C} \textbf{\bibinfo{volume}{61}},
  \bibinfo{pages}{024319} (\bibinfo{year}{2000}).

\bibitem[{\citenamefont{Skaza et~al.}(2005)}]{Sk05}
\bibinfo{author}{\bibfnamefont{F.}~\bibnamefont{Skaza}} \bibnamefont{et~al.},
  \bibinfo{journal}{Phys. Lett.} \textbf{\bibinfo{volume}{B619}},
  \bibinfo{pages}{82} (\bibinfo{year}{2005}).

\bibitem[{\citenamefont{Zheng et~al.}(1995)\citenamefont{Zheng, Barrett, Vary,
  Haxton, and Song}}]{Zh95}
\bibinfo{author}{\bibfnamefont{D.~C.} \bibnamefont{Zheng}},
  \bibinfo{author}{\bibfnamefont{B.~R.} \bibnamefont{Barrett}},
  \bibinfo{author}{\bibfnamefont{J.~P.} \bibnamefont{Vary}},
  \bibinfo{author}{\bibfnamefont{W.~C.} \bibnamefont{Haxton}},
  \bibnamefont{and} \bibinfo{author}{\bibfnamefont{C.-L.} \bibnamefont{Song}},
  \bibinfo{journal}{Phys. Rev. C} \textbf{\bibinfo{volume}{52}},
  \bibinfo{pages}{2498} (\bibinfo{year}{1995}).

\bibitem[{\citenamefont{Tilley et~al.}(2002)\citenamefont{Tilley, Cheves,
  Godwin, Hale, Hofmann, Kelly, Sheu, and Weller}}]{TUNL}
\bibinfo{author}{\bibfnamefont{D.~R.} \bibnamefont{Tilley}},
  \bibinfo{author}{\bibfnamefont{C.~M.} \bibnamefont{Cheves}},
  \bibinfo{author}{\bibfnamefont{J.~L.} \bibnamefont{Godwin}},
  \bibinfo{author}{\bibfnamefont{G.~M.} \bibnamefont{Hale}},
  \bibinfo{author}{\bibfnamefont{H.~M.} \bibnamefont{Hofmann}},
  \bibinfo{author}{\bibfnamefont{J.~H.} \bibnamefont{Kelly}},
  \bibinfo{author}{\bibfnamefont{C.~G.} \bibnamefont{Sheu}}, \bibnamefont{and}
  \bibinfo{author}{\bibfnamefont{H.~R.} \bibnamefont{Weller}},
  \bibinfo{journal}{Nucl. Phys.} \textbf{\bibinfo{volume}{A708}},
  \bibinfo{pages}{3} (\bibinfo{year}{2002}).

\bibitem[{\citenamefont{Jouanne et~al.}(2005)}]{Jo05}
\bibinfo{author}{\bibfnamefont{C.}~\bibnamefont{Jouanne}} \bibnamefont{et~al.},
  \bibinfo{journal}{Phys. Rev. C} \textbf{\bibinfo{volume}{72}},
  \bibinfo{pages}{014308} (\bibinfo{year}{2005}).

\bibitem[{\citenamefont{Warburton and Millener}(1989)}]{Wa89}
\bibinfo{author}{\bibfnamefont{E.~K.} \bibnamefont{Warburton}}
  \bibnamefont{and} \bibinfo{author}{\bibfnamefont{D.~J.}
  \bibnamefont{Millener}}, \bibinfo{journal}{Phys. Rev. C}
  \textbf{\bibinfo{volume}{39}}, \bibinfo{pages}{1120} (\bibinfo{year}{1989}).

\bibitem[{\citenamefont{Warburton and Brown}(1992)}]{Wa92}
\bibinfo{author}{\bibfnamefont{E.~K.} \bibnamefont{Warburton}}
  \bibnamefont{and} \bibinfo{author}{\bibfnamefont{B.~A.} \bibnamefont{Brown}},
  \bibinfo{journal}{Phys. Rev. C} \textbf{\bibinfo{volume}{46}},
  \bibinfo{pages}{923} (\bibinfo{year}{1992}).

\bibitem[{\citenamefont{McCarthy et~al.}(1977)\citenamefont{McCarthy, Sick, and
  Whitney}}]{Mc77}
\bibinfo{author}{\bibfnamefont{J.~S.} \bibnamefont{McCarthy}},
  \bibinfo{author}{\bibfnamefont{I.}~\bibnamefont{Sick}}, \bibnamefont{and}
  \bibinfo{author}{\bibfnamefont{R.~R.} \bibnamefont{Whitney}},
  \bibinfo{journal}{Phys. Rev. C} \textbf{\bibinfo{volume}{15}},
  \bibinfo{pages}{1396} (\bibinfo{year}{1977}).

\bibitem[{\citenamefont{Ottermann et~al.}(1985)\citenamefont{Ottermann,
  K{\"o}bschall, Maurer, R{\"o}rich, Schmitt, and Walter}}]{Ot85}
\bibinfo{author}{\bibfnamefont{C.~R.} \bibnamefont{Ottermann}},
  \bibinfo{author}{\bibfnamefont{G.}~\bibnamefont{K{\"o}bschall}},
  \bibinfo{author}{\bibfnamefont{K.}~\bibnamefont{Maurer}},
  \bibinfo{author}{\bibfnamefont{K.}~\bibnamefont{R{\"o}rich}},
  \bibinfo{author}{\bibfnamefont{C.}~\bibnamefont{Schmitt}}, \bibnamefont{and}
  \bibinfo{author}{\bibfnamefont{V.~H.} \bibnamefont{Walter}},
  \bibinfo{journal}{Nucl. Phys.} \textbf{\bibinfo{volume}{A436}},
  \bibinfo{pages}{688} (\bibinfo{year}{1985}).

\bibitem[{\citenamefont{Suelzle et~al.}(1967)\citenamefont{Suelzle, Yearian,
  and Crannell}}]{Su67}
\bibinfo{author}{\bibfnamefont{L.~R.} \bibnamefont{Suelzle}},
  \bibinfo{author}{\bibfnamefont{M.~R.} \bibnamefont{Yearian}},
  \bibnamefont{and} \bibinfo{author}{\bibfnamefont{H.}~\bibnamefont{Crannell}},
  \bibinfo{journal}{Phys. Rev.} \textbf{\bibinfo{volume}{162}},
  \bibinfo{pages}{992} (\bibinfo{year}{1967}).

\bibitem[{\citenamefont{Lichtenstadt et~al.}(1989)\citenamefont{Lichtenstadt,
  Alster, Moinester, Dubach, Hicks, Peterson, and Kowlaski}}]{Li89}
\bibinfo{author}{\bibfnamefont{J.}~\bibnamefont{Lichtenstadt}},
  \bibinfo{author}{\bibfnamefont{J.}~\bibnamefont{Alster}},
  \bibinfo{author}{\bibfnamefont{M.~A.} \bibnamefont{Moinester}},
  \bibinfo{author}{\bibfnamefont{J.}~\bibnamefont{Dubach}},
  \bibinfo{author}{\bibfnamefont{R.~S.} \bibnamefont{Hicks}},
  \bibinfo{author}{\bibfnamefont{G.~A.} \bibnamefont{Peterson}},
  \bibnamefont{and} \bibinfo{author}{\bibfnamefont{S.}~\bibnamefont{Kowlaski}},
  \bibinfo{journal}{Phys. Lett.} \textbf{\bibinfo{volume}{B219}},
  \bibinfo{pages}{394} (\bibinfo{year}{1989}).

\bibitem[{\citenamefont{van Niftrik et~al.}(1971)\citenamefont{van Niftrik,
  Lapik\'as, de~Vries, and Box}}]{Ni71}
\bibinfo{author}{\bibfnamefont{G.~J.~C.} \bibnamefont{van Niftrik}},
  \bibinfo{author}{\bibfnamefont{L.}~\bibnamefont{Lapik\'as}},
  \bibinfo{author}{\bibfnamefont{H.}~\bibnamefont{de~Vries}}, \bibnamefont{and}
  \bibinfo{author}{\bibfnamefont{G.}~\bibnamefont{Box}},
  \bibinfo{journal}{Nucl. Phys.} \textbf{\bibinfo{volume}{A174}},
  \bibinfo{pages}{173} (\bibinfo{year}{1971}).

\bibitem[{\citenamefont{Tilley et~al.}(2004)\citenamefont{Tilley, Kelley,
  Godwin, Millener, Purcell, Sheu, and Weller}}]{TUNL04}
\bibinfo{author}{\bibfnamefont{D.~R.} \bibnamefont{Tilley}},
  \bibinfo{author}{\bibfnamefont{J.~H.} \bibnamefont{Kelley}},
  \bibinfo{author}{\bibfnamefont{J.~L.} \bibnamefont{Godwin}},
  \bibinfo{author}{\bibfnamefont{D.~J.} \bibnamefont{Millener}},
  \bibinfo{author}{\bibfnamefont{J.~E.} \bibnamefont{Purcell}},
  \bibinfo{author}{\bibfnamefont{C.~G.} \bibnamefont{Sheu}}, \bibnamefont{and}
  \bibinfo{author}{\bibfnamefont{H.~R.} \bibnamefont{Weller}},
  \bibinfo{journal}{Nucl. Phys.} \textbf{\bibinfo{volume}{A745}},
  \bibinfo{pages}{155} (\bibinfo{year}{2004}).

\end{thebibliography}

\end{document}